\documentclass[a4paper,12pt]{article}
\usepackage[pctex32]{graphics}
\usepackage{amssymb,amsmath}

\begin{document}

\topmargin 0pt
\oddsidemargin 0mm
\newcommand{\be}{\begin{equation}}
\newcommand{\ee}{\end{equation}}
\newcommand{\ba}{\begin{eqnarray}}
\newcommand{\ea}{\end{eqnarray}}
\newcommand{\fr}{\frac}

\renewcommand{\thefootnote}{\fnsymbol{footnote}}

\begin{titlepage}

\vspace{5mm}
\begin{center}
{\Large \bf Entropy of extremal warped black holes}

\vskip .6cm
 \centerline{\large Yun Soo Myung$~^a$}

\vskip .6cm

{Institute of Basic Science and School of Computer Aided
Science,
\\Inje University, Gimhae 621-749, Korea }

\end{center}

\vspace{5mm}
 \centerline{{\bf{Abstract}}}

\vspace{5mm}

We study the entropy of extremal warped black hole obtained from
the topologically massive gravity with a negative cosmological
constant of $\Lambda=-1/l^2$. We compare the entropy $S_e=\pi
\alpha/3G$ from the Wald formalism with  $S_w=\pi l u /3G$ from
the entropy function approach. These are the same if $\alpha=l u$.
Also we obtain the same Cardy formula when $J_e= l^3 q$ with $J_e$
the angular momentum and $q$ the conserved quantity.

\vspace{5mm}


\vskip 0.8cm

\noindent $^a$ysmyung@inje.ac.kr

\noindent
\end{titlepage}

\newpage

\renewcommand{\thefootnote}{\arabic{footnote}}
\setcounter{footnote}{0} \setcounter{page}{2}

\section{Introduction}
The gravitational Chern-Simons terms with $K$ the gravitational
Chern-Simons coupling constant produces a physically propagating
massive graviton  in three dimensional Einstein
gravity~\cite{DJT}. This topologically massive gravity with a
negative cosmological constant $\Lambda=-1/l^2$ (TMG$_\Lambda$)
gives us the BTZ black hole solution~\cite{LSS1,LSS2}. For
$l/K>3(\nu>1)$, there exists warped black hole solutions which are
asymptotic to warped AdS$_3$ spacetimes~\cite{ALPSS}. These warped
black holes are considered discrete quotients by an element of
$SL(2,R)\times U(1)$ of warped AdS$_3$, as the BTZ black holes are
discrete quotients of AdS$_3$. Although thermodynamic quantities
including the entropy have been investigated,  their forms are
very complicated. Especially, the entropy is still not fully
understood~\cite{Ann,KS}.

We note that the gravitational Chern-Simons terms are not
invariant under coordinate transforms even they are conformally
invariant~\cite{GIJP,GK}. Thus, their variation of Cotten tensor
plays no role in finding a new solution.  All solutions of
Einstein gravity are solutions of the TMG$_\Lambda$. Hence, one
needs to seek  an another way to investigate the TMG$_\Lambda$. In
this end, one may introduce conformal transformation to single out
a conformal degree of freedom (dilaton).  Then, the Kaluza-Klein
ansatz is used to obtain an effective two-dimensional action of
2DTMG$_\Lambda$, which will be a gauge and coordinate invariant.
Saboo and Sen~\cite{SSen} have used the 2DTMG$_\Lambda$ to obtain
the entropy of extremal BTZ black hole by using the entropy
function formalism. This is possible because AdS$_2$ is stable
attractor solution of equations which govern how the geometry
changes as the degenerate horizon is approached. On the other
hand, for a constant dilaton, the authors in~\cite{AFM,MKP,KMP}
have employed  the entropy function approach to find three
distinct vacuum solutions of the 2DTMG$_\Lambda$: AdS$_2$ with
positive charge, AdS$_2$ with negative charge, and warped AdS$_2$
with positive charge. Upon uplifting  to three dimensions, these
were geometric solutions which are either AdS$_3$ or warped
AdS$_3$ with an identification.

In this work, we explore the connection between  two entropies of
the extremal warped black holes obtained using  the  Wald
formalism and the entropy function method.

\section{Topologically massive gravity in Schwarzschild coordinates}
The action for  topologically massive gravity with a negative
cosmological constant (TMG$_\Lambda$) is given
by~\cite{DJT,CDWW1,GJ,GKP,MyungL,Park,GJJ,CDWW2,Carl,Stro,BC}
\begin{equation} \label{tmg}
I_{\rm TMG_\Lambda}=\frac{1}{16 \pi G_3}\int
d^3x\sqrt{-g}\Bigg[R_3 -2\Lambda +
\frac{K}{2}\varepsilon^{lmn}\Gamma^p_{~lq}
\Big(\partial_{m}\Gamma^q_{~np}+\frac{2}{3}\Gamma^q_{~mr}\Gamma^r_{~np}\Big)\Bigg],
\end{equation}
where $\varepsilon$ is the tensor  defined by $\epsilon/\sqrt{-g}$
with $\epsilon^{012}=1$. We choose the Newton's constant $G_3>0$.
The Latin indices of $l,m,n, \cdots$ denote three dimensional
tensors. The $K(=1/\mu)$-term is called the gravitational
Chern-Simons terms. Here we choose ``+" sign to avoid negative
graviton energy~\cite{GJ}. It is the first higher derivative
correction in three dimensions because it is the third-order
derivative.

Varying this action leads to the Einstein equation
\begin{equation} \label{eineq}
G_{mn} + KC_{mn}=0,
\end{equation}
where the Einstein tensor including the cosmological constant is
given by
\begin{equation}
G_{mn}=R_{3mn}-\frac{R_3}{2}g_{mn}
-\frac{1}{l^2}g_{mn}
\end{equation}
and the Cotton tensor is
\begin{equation}
C_{mn}= \varepsilon_m~^{pq}\nabla_p
\Big(R_{3qn}-\frac{1}{4}g_{qn}R_3\Big).
\end{equation}
We note that the Cotton tensor $C_{mn}$ vanishes for
 any solution to Einstein gravity, so all solutions to general
 relativity are also solutions of the TMG$_\Lambda$. Hence, for $K=0$, the BTZ black hole solution~\cite{BTZ}
  is obtained as the vacuum solution to the Einstein gravity with
  $K=0~(G_{mn}=0$),
\begin{equation}
ds^2_{BTZ} = -N^2(r) dt^2 + \frac{dr^2}{N^2(r)} + r^2 \Big(d\theta
+ N^\theta(r) dt\Big)^2 , \label{btzmetric}
\end{equation}
where the metric function $N^2(r)$ and the lapse function
$N^\theta(r)$ are given by
\begin{equation}
N^2(r) = -8 G_3 m + \frac{r^2}{l^2} + \frac{16 G_3^2 j^2}{r^2},~~
N^\theta(r) = - \frac{4 G_3 j}{r^2}.\label{def_N}
\end{equation}
 Here $m$ and $j$ are the mass and angular momentum of the BTZ
 black hole, respectively.

On the other hand, for $\nu^2=(l/3K)^2>1$,  the warped black hole
solution
 is obtained as the vacuum solution to the
TMG$_\Lambda$ \cite{ALPSS,CD,Ann,OK,KS,CX,NP}
\begin{equation}
ds^2_{wBH}=-\tilde{N}^2dt^2+\frac{l^4}{4\tilde{R}^2\tilde{N}^2}dr^2+l^2\tilde{R}^2\Big(d\theta+\tilde{N}^\theta
dt\Big)^2,
\end{equation}
where
\begin{eqnarray}
 && \tilde{R}^2(r)=\frac{r}{4}\left(3(\nu^2-1)r +(\nu^2+3)(r_++r_-)-4\nu\sqrt{r_+r_-(\nu^2+3)}\right),\nonumber\\
 && \tilde{N}^2(r)=\frac{l^2(\nu^2+3)(r-r_+)(r-r_-)}{4\tilde{R}^2(r)},\nonumber\\
 && \tilde{N}^\theta(r)=\frac{2\nu r-\sqrt{r_+r_-(\nu^2+3)}}{2\tilde{R}^2(r)}.
\end{eqnarray}
We note that this warped black hole   reduces to the BTZ black
hole in a rotating frame when choosing $\nu^2=1$. Using the
surface integral expressions, thermodynamic quantities of the ADT
mass $M$ and angular momentum $J$ are obtained as
\begin{eqnarray}
 && M=\frac{(\nu^2+3)}{24G}\Big(r_++r_--\frac{1}{\nu}\sqrt{r_+r_-(\nu^2+3)}\Big),\\
 && J=\frac{\nu l(\nu^2+3)}{96G}\Bigg[\Big(r_++r_--\frac{1}{\nu}\sqrt{r_+r_-(\nu^2+3)}\Big)^2
  -\frac{(5\nu^2+3)}{4\nu^2}(r_+-r_-)^2\Bigg].
\end{eqnarray}
We note the length dimensions of $M$ and $J$: $[M]=0,~[J]=2$ with
$[G]=[l]=[K]=1$. It was conjectured that the warped black holes
are holographically dual to a two-dimensional conformal field
theory (CFT$_2$). The study of thermodynamics of these black holes
have provided a strong support on the AdS$_3$/CFT$_2$
correspondence. Actually, the left  sector  of the CFT$_2$ is
independent of the right sector. At the thermal equilibrium, the
two sectors have different temperatures: right and left moving
temperatures given by
\begin{eqnarray}
 && T_R=\frac{(\nu^2+3)(r_+-r_-)}{8 \pi l},\\
 && T_L=\frac{(\nu^2+3)}{8 \pi l}\Bigg[r_++r_--\frac{1}{\nu}\sqrt{r_+r_-(\nu^2+3)}\Bigg].
\end{eqnarray}
Using these, one defines the Hawking temperature $T_H$
\begin{equation}
\frac{1}{T_H}=\frac{4 \pi \nu l}{\nu^2+3}\frac{T_L+T_R}{T_R}.
\end{equation}
The entropy  is calculated by using the Wald method which is a
prescription for handling  higher derivative terms. It is composed
of two terms from  Einstein action and Chern-Simons action as
\begin{equation}
S=\frac{\pi }{24 \nu G}\Big[(9\nu^2+3)r_+-(\nu^2+3)r_--4\nu\sqrt{r_+r_-(\nu^2+3)}\Big].
 \end{equation}
Defining right and left moving charges
\begin{eqnarray}
\label{cent-r} && c_R=\frac{(5\nu^2+3)l}{G\nu (\nu^2+3)},\\
\label{cent-l} && c_L=\frac{4\nu l}{G (\nu^2+3)},
\end{eqnarray}
the above entropy can be rewritten as
\begin{equation}
S=\frac{\pi^2 }{3}\Big(c_LT_L+c_RT_R\Big).
 \end{equation}
This is the formula for the entropy of CFT$_2$ with central charges $c_L$ and $c_R$
at temperatures $T_L$ and $T_R$.
Introducing right and left moving energies
\begin{equation}
E_R=\frac{\pi^2l c_R T_R^2}{6},~~E_L=\frac{\pi^2l c_L T_L^2}{6},
 \end{equation}
 one expresses the angular momentum in terms of a difference of these energies
 \begin{equation}
J=l\Big(E_L-E_R\Big).
 \end{equation}
Here we observe the length dimensions of
$[E_R]=[E_L]=1,~[T_R]=[T_L]=0,~[c_R]=[c_L]=0,~[S]=0$. Furthermore,
one could express the entropy as sum of the square roots of left
and right energies
  \begin{equation}
S=2 \pi \Bigg( \sqrt{\frac{E_R}{6l}~c_L}+\sqrt{\frac{E_R}{6l}~c_L}\Bigg),
 \end{equation}
 which seems to be a Cardy formula for CFT$_2$.
 In this sense, the conserved charges ($E_L,E_R$) and potentials ($T_L,T_R$) are more natural
 for describing the warped black holes than ($M,J$) and ($T_H,\Omega_H$) with $\Omega_H=
 2/(2\nu r_+-\sqrt{(\nu^2+3)r_+r_-})l$ the angular velocity of the horizon.

 In this work, this formalism is  attractive because we consider the extremal
 warped black hole  with $T_R=0(E_R=0)$ at $r_+=r_-\equiv r_e$.
 In this case, the relevant quantities are angular momentum and left moving  temperature
 \begin{equation}
  J_e=\frac{\nu l(\nu^2+3)r_e^2}{96G}\Bigg(2-\frac{\sqrt{\nu^2+3}}{\nu}\Bigg)^2,~~
  T^e_L=\frac{(\nu^2+3)r_e}{8 \pi l}\Bigg(2-\frac{\sqrt{\nu^2+3}}{\nu}\Bigg).
\end{equation}
Importantly,  the entropy for extremal warped black hole takes the form
 \begin{equation}\label{eent}
S_e=\frac{\pi  \nu r_e}{6G}\Bigg(2-\frac{\sqrt{\nu^2+3}}{\nu}\Bigg).
 \end{equation}
 Here we observe that the positive entropy is guaranteed only for
 $\nu>1$.
Also, $S_e$ could be  recovered from the relation of $S_e=\pi^2
c_LT^e_L/3$. At this stage, since we do not know the location
$r_e$ of extremal black hole explicitly, we could eliminate $r_e$
by rewriting $S_e$ in terms of $J_e$ and $c_L$ as
\begin{equation} \label{cardy-f}
S_e= 2 \pi \sqrt{\frac{J_e}{6l^2}~c_L}=2 \pi
\sqrt{\frac{E^e_L}{6l}~c_L}
 \end{equation}
 which looks like the Cardy formula for left mover.
 Here we used a relation of $J_e=l E_L^e$, which shows the
 extremality of warped black hole.
 However, the relation between $S_e$ and $M_e$ is not
 what one really wants to find like the Cardy formula
  \begin{equation}
S_e=\Bigg(\frac{4\pi  \nu }{\nu^2+3}\Bigg)M_e,
 \end{equation}
 with
  \begin{equation}
M_e=\frac{(\nu^2+3) r_e}{24G}\Bigg(2-\frac{\sqrt{\nu^2+3}}{\nu}\Bigg).
 \end{equation}
It shows  a linear relation  between $S_e$ and $M_e$. Another
interesting relation  is
\begin{equation}
M_e=\Bigg[\frac{\pi l}{3G}\Bigg]T^e_L.
\end{equation}
Finally, the first law of thermodynamics seems to be
\begin{equation}
d\Big(\frac{E^e_L}{l}\Big)= T^e_L dS_e.
\end{equation}

We note that the above description of extremal warped black hole
might not be the only type of geometry which should be considered
as a black hole. The boundary of the Poincare patch of AdS$_2$ or
AdS$_3$ is an event horizon when these spaces arise as
near-horizon limits of extremal warped black holes. There is a
Killing horizon associated to time translations but no singularity
befind it. In the next section, we will study the extremal warped
black hole using  Poincare coordinates.

\section{Entropy function approach in Poincare coordinates}
We first make  conformal transformation and then perform
dimensional reduction using the metric~\cite{GIJP,GK}
\begin{equation} \label{2Dmetric}
ds^2_{DR}=\phi^2\Big[g_{\mu\nu}(x)dx^\mu dx^\nu+(dy+A_\mu
(x)dx^\mu)^2\Big],
\end{equation}
where $y$ is coordinate that parameterizes an $S^1$ with period $2
\pi l$. Hence, its isometry is factorized as ${\cal G}\times
U(1)$. After ``$y$"-integration, the action (\ref{tmg}) reduces to
the effective two-dimensional action of
2DTMG$_\Lambda$~\cite{AFM,MKP,KMP}
\begin{eqnarray}
 I_{\rm 2DTMG_\Lambda}&=& \frac{l}{8G_3}\int d^2x \sqrt{-g}\left(\phi R+\frac{2}{\phi}g^{\mu\nu}\nabla_\mu\phi\nabla_\nu\phi
     +\frac{2}{l^2}\phi^3-\frac{1}{4}\phi F_{\mu\nu}F^{\mu\nu}\right)\nonumber\\
 &+& \frac{Kl}{32 G_3} \int d^2x \left(R\epsilon^{\mu\nu}F_{\mu\nu}+
     \epsilon^{\mu\nu}F_{\mu\rho}F^{\rho\sigma}F_{\sigma\nu}\right).
\end{eqnarray}
Here $R$ is the 2D Ricci scalar with $R_{\mu\nu}=Rg_{\mu\nu}/2$
and $\phi$ is the dilaton. Also
$F_{\mu\nu}=\partial_{\mu}A_{\nu}-\partial_{\nu}A_{\mu}$ is the
field strength for gravivector $A_\mu$ with $\epsilon^{01}=1$. The
Greek indices of $\mu,\nu, \rho, \cdots$ represent two dimensional
tensors.  We note that this action was used to derive the entropy
of extremal BTZ black hole when applying the entropy function
approach~\cite{SSen,AFM}.

Introducing  a dual notation of $F$ through
$F^{\mu\nu}=\frac{1}{\sqrt{-g}}\epsilon^{\mu\nu}F$, equations of
motion for dilaton  $\phi$ and gravivector $A_\mu$, respectively,
are given by
\begin{eqnarray}
 &&
  \label{EOM-phi}
 R+\frac{2}{\phi^2}(\nabla\phi)^2-\frac{4}{\phi}\nabla^2\phi+\frac{6}{l^2}\phi^2+\frac{1}{4}F^2=0,\\
 && \label{EOM-A}
  \epsilon^{\mu\nu}\partial_\nu\left[\phi
  F-\frac{K}{2}(R+3F^2)\right]=0.
\end{eqnarray}
The trace part of Einstein equation
\begin{equation}
 \label{trgEOM}
 \nabla^2\phi-\frac{2}{l^2}\phi^3+\frac{1}{2}\phi F^2
 -K\left(\frac{1}{2}RF+F^3+\frac{1}{2}\nabla^2F\right)=0
\end{equation}
 is relevant to the entropy calculation.
Eq.(\ref{EOM-A}) implies
\begin{equation}
 \label{trlessgEOM}
 \phi F-\frac{K}{2}(R+3F^2)={\cal C}
\end{equation}
where ${\cal C}$ is a constant related to a conserved quantity.
Now, we wish to find AdS$_2$ as the vacuum solution to Eqs.
(\ref{EOM-phi}) and (\ref{trgEOM}). In the case of a constant
dilaton, we obtain  the condition from these equations,
\begin{equation}
(3KF-2\phi)\left(\frac{\phi^2}{l^2}-\frac{1}{4}F^2\right)=0,
\end{equation}
which implies three relations between $\phi$ and $F$
\begin{eqnarray}
\label{ads}
&& \phi_{\pm} = \pm\frac{l}{2}F, \\
\label{warp} && \phi_{w} = \frac{3K}{2}F.
\end{eqnarray}
 $\phi_\pm=\pm lF/2$
denote the vacuum solutions to the 3D Einstein gravity.  Hereafter
we consider the case of (\ref{warp})  only because we are
interested in the vacuum solution to the 2DTMG$_\Lambda$.
 We note that for the case of $K=l/3$, $\phi_w=\phi_+$
which is a degenerate vacuum.

Assuming the line element preserving ${\cal G}=SL(2,R)$ isometry
\begin{equation}
ds^2_{AdS_2}=v\left(-x^2d\psi^2+\frac{dx^2}{x^2}\right),
\end{equation}
we have the AdS$_2$-spacetimes, which satisfies
\begin{equation} \label{ads2}
\bar{R}=-\frac{2}{v},
~~~\bar{\phi}=u,~~~\bar{F}=e/v~(\bar{F}_{10}=e).
\end{equation}
In order to find an explicit form of the solution, we have to use
the entropy function formalism because it provides an efficient
way to find warped  AdS$_2$-solution as well as entropy of
extremal warped black hole.  The entropy function is defined as
\begin{equation} \label{entf}
{\cal E}=2\pi\Big(qe-\tilde{f}(e,v,u)\Big)
\end{equation}
where $\tilde{f}(e,v,u)$ is the Lagrangian density ${\cal L}_{\rm
2DTMG_\Lambda}$ evaluated when using Eq. (\ref{ads2}),
\begin{equation}
 \tilde{f}=\frac{l}{8G}\Bigg[-2u+\frac{2u^3v}{l^2}+\frac{ue^2}{2v}
   +\frac{K}{2}\left(\frac{2e}{v}-\frac{e^3}{v^2}\right)\Bigg].
\end{equation}
Here we have equations of motion upon the variation of ${\cal E}$
with respect to $u$, $v$, and $e$
\begin{eqnarray}
 \label{fequ}&& -2+\frac{6u^2v}{l^2}+\frac{e^2}{2v}=0,\\
 \label{feqv}&& \frac{2u^3}{l^2}-\frac{ue^2}{2v^2}
    -K\left(\frac{e}{v^2}-\frac{e^3}{v^3}\right)=0,\\
\label{feqe} &&
q=\frac{l}{8G}\Bigg[\frac{ue}{v}-\frac{K}{2}\left(\frac{2}{v}-\frac{3e^2}{v^2}\right)\Bigg].
\end{eqnarray}
For a consistency check, we mention the length dimensions of
$[v]=2,~[u]=0,~[e]=1,~[q]=-1$. Equations (\ref{fequ}) and
(\ref{feqv}) are those obtained by plugging Eq. (\ref{ads2}) into
Eqs. (\ref{EOM-phi}) and (\ref{trgEOM}). This means that the
entropy function formalism uses the Einstein equation and dilaton
equation in  near-horizon geometry  AdS$_2 \times S^1$ of the
extremal warped  black hole. The difference is that Eq.
(\ref{EOM-A}) is trivially satisfied with Eq. (\ref{ads2}), while
Eq. (\ref{feqe}) is useful for deriving the entropy by choosing
$q=\frac{l{\cal C}}{8G}$. The conserved quantity $q$ measures
momentum along $y$, which for warped black hole, is closely
related to the angular momentum $J_e$~\cite{SSen}.

Here we obtain the warped AdS$_2$-solution  for
$u=3Ke/2v(\phi_w=3FK/2)$,
\begin{equation}
u=\sqrt{\frac{72GKql}{l^2+27K^2}},~~~v=\frac{Kl}{8Gq},~~~e=\sqrt{\frac{Kl^3}{2Gq(l^2+27K^2)}},~~~q>0.
\end{equation}
Then, plugging these into Eq.(\ref{entf}) leads to the  entropy of
the extremal warped black hole
\begin{equation} \label{DRent}
 S_w= \frac{2\pi}{eG} \frac{Kl^3}{l^2+27K^2}=4\pi eq \simeq 2\pi\sqrt{\frac{ql}{6}
 c_L},~~~K>0,
\end{equation}
where the left moving central charge (\ref{cent-l}) takes the form
\begin{equation}
c_L=\frac{24 q e^2 }{l}
\end{equation}
when employing the twisted energy momentum tensor for the
$k=8ql^2$ level of $U(1)$ current with $c_L=3k
e^2/l^4$~\cite{AFM}.
 We check  the correct length dimension  $[S_w]=0$ of the entropy  together with $[ql]=0$ and $[c_L]=0$.
The last relation ($\simeq$) will be confirmed from the Cardy
formula if $ql$ is the eigenvalue of $L_0$-operator of dual
CFT$_2$. If one assumes the relation
\begin{equation}
J_e=l^3 q,
\end{equation}
one recovers the Cardy formula (\ref{cardy-f}). We note that the
background metric (\ref{2Dmetric}) can be rewritten as the
extremal warped  black hole in Poincare coordinates ($\psi, x,
z$)~\cite{ALPSS}
\begin{eqnarray}
ds^2_{DR}&=&(\bar{\phi})^2\Big[\bar{g}_{\mu\nu}dx^\mu
dx^\nu+(dy+\bar{A}_\mu
dx^\mu)^2\Big]\\
&=& u^2 v\Big(-x^2d\psi^2+\frac{dx^2}{x^2}\Big)+(ue)^2\Big(dz^2+xd\psi\Big)^2 \\
&=& \frac{l^2}{\nu^2+3}\Bigg[-x^2d\psi^2+\frac{dx^2}{x^2}+\frac{4\nu^2}{\nu^2+3}(dz+xd\psi)^2 \Bigg]
\end{eqnarray}
with $y=ez$. This appears to be independent of $q$.

 Finally, we express the extremal entropy $S_w$
in Eq.(\ref{DRent}) in terms of Poincare coordinates as
\begin{equation}
S_w=\frac{\pi l u}{3G},~~u=\bar{\phi}=\sqrt{\frac{24G\nu
q}{\nu^2+3}}.
\end{equation}
In Schwarzschild coordinates, self-dual solutions are obtained by
identifying time $t$, i.e. $t \sim t+2\pi \alpha$. The left moving
temperature is given by $T_L=(\nu^2+3)\alpha/(4\pi \nu l )$. Then,
the entropy (\ref{eent}) takes the form
\begin{equation}
S_{e}=\frac{\pi \alpha}{3G}
\end{equation}
with
\begin{equation}
\alpha=\frac{\nu r_e}{2} \Bigg(2-\frac{\sqrt{\nu^2+3}}{\nu}\Bigg).
\end{equation}
Since the entropy is an invariant quantity, comparing  $S_w$ with
$S_e$ leads to the important relation
\begin{equation} \label{alphau}
\alpha=l ~u.
\end{equation}
We may understand the relation between $S_{e}$ and $S_w$ by considering the coordinate transformations
\begin{eqnarray}
 \label{cordtt}t&=& -\frac{2\alpha}{\nu^2+3} \psi +\frac{2\nu}{\nu^2+3}z, \\
 \label{cordth} \theta&=& \frac{2}{\nu^2+3}\psi,\\
 \label{cordrx} r&=&x+r_e.
 \end{eqnarray}
The location  $r=r_e$ of extremal warped black hole appears at the
boundary of $x=0$. We observe ``$\alpha$" in Eq.(\ref{cordtt}),
even though there does not exist  the exact relation of
Eq.(\ref{alphau}).

In conclusion,  in order to have the same entropy for extremal
warped black hole, we observe the correspondences between two
approaches as
\begin{equation}
\alpha \longleftrightarrow  l ~u,~~~ J_e \longleftrightarrow l^3
~q,
\end{equation}
where the latter is introduced  for obtaining  the same Cardy
formula. The difference is that in the Wald formalism with
Schwarzschild coordinates, the positive entropy is allowed for
$\nu>1$ and the extremal entropy is problematic at $\nu=1$, while
in the entropy function formalism with Poincare coordinates, the
positive entropy is guaranteed for $\nu>0$ and the extremal
entropy is properly obtained for $\nu=1$.

\section*{Acknowledgement}
The author thanks Y.-J. Park and Y.-W. Kim  for helpful discussions. This work  was supported by the Korea Research Foundation (KRF-2006-311-C00249)
funded by the Korea Government (MOEHRD).

\end{document}